\documentclass[]{aa}

\usepackage{graphicx}
\usepackage{txfonts}
\usepackage{natbib}

\newcommand{\un}[1]{{\,{\rm #1}}}
\newcommand{\dw}[1]{_{\scriptstyle \mathrm{#1}}}
\newcommand{\up}[1]{^{\scriptstyle \mathrm{#1}}}

\begin{document}

\title{A further `degree of freedom' in the rotational evolution of
stars}

\author{V Holzwarth \and M Jardine}
\offprints{V Holzwarth, \email{vrh1@st-and.ac.uk}}

\institute{School of Physics and Astronomy, University of St Andrews, 
North Haugh, St Andrews, Fife KY16 9SS, Scotland}

\date{Received ; accepted}

\abstract{
Observational and theoretical investigations provide evidence for
non-uniform spot and magnetic flux distributions on rapidly rotating
stars, which have a significant impact on their angular momentum loss
rate through magnetised winds.
Supplementing the formalism of \citet{1991ApJ...376..204M} with a
latitude-dependent magnetised wind model, we analyse the effect of
analytically prescribed surface distributions of open magnetic flux
with different shapes and degrees of non-uniformity on the rotational
evolution of a solar-like star.
The angular momentum redistribution inside the star is treated in a
qualitative way, assuming an angular momentum transfer between the
rigidly-rotating radiative and convective zones on a constant coupling
timescale of $15\un{Myr}$; for the sake of simplicity we disregard
interactions with circumstellar disks.
We find that non-uniform flux distributions entail rotational histories
which differ significantly from those of classical approaches, with
differences cumulating up to $200\%$ during the main sequence phase.
Their impact is able to mimic deviations of the dynamo efficiency from
linearity of up to $40\%$ and nominal dynamo saturation limits at about
$35$ times the solar rotation rate.
Concentrations of open magnetic flux at high latitudes thus assist in
the formation of very rapidly rotating stars in young open clusters,
and ease the necessity for a dynamo saturation at small rotation
rates.
However, since our results show that even minor amounts of open flux at
intermediate latitudes, as observed with Zeeman-Doppler imaging
techniques, are sufficient to moderate this reduction of the AM loss
rate, we suggest that non-uniform flux distributions are a
complementary rather than an alternative explanation for very rapid
stellar rotation.
\keywords{Stars: rotation -- Stars: winds, outflows -- Stars: magnetic
fields -- Stars: mass-loss -- Stars: evolution -- MHD}
}

\maketitle

\section{Introduction}
\label{intro}
In the presence of open magnetic fields the angular momentum (AM) loss
of a star through winds and outflows is significantly enhanced since
the tension of bent field lines effectively prolongs the lever arm of
the associated torque \citep{1962AnAp...25...18S}.
Whereas during the pre-main sequence (PMS) phase the rotational
evolution of a star is dominated by its changing stellar structure and
magnetic interaction with a circumstellar accretion disk, its rotation
during the main sequence (MS) phase is mainly determined through
braking by magnetised winds \citep{1976ApJ...210..498B}.
Theoretical studies of magnetised winds go back to classical approaches
of \citet[ hereafter WD]{1967ApJ...148..217W} and
\citet{1968MNRAS.138..359M}.
At high rotation rates the magnetic field adds considerably to the
acceleration of the outflow through magneto-centrifugal driving
\citep{1969ApJ...158..727M}, which makes magnetised winds intrinsically
latitude-dependent.
The total AM loss rate is consequently susceptible to variations of the
surface magnetic field \citep{1997A&A...325.1039S, 2005smaamldtlmw} and
atmospheric field topology \citep{1987MNRAS.226...57M,
1988ApJ...333..236K}.
More recently, multi-dimensional MHD-simulations have been accomplished
to study structural and temporal wind properties like coronal mass
ejections in more detail \citep[e.g.,][]{1985A&A...152..121S,
1999A&A...343..251K, 2000ApJ...530.1036K}; the extensive computational
requirements render them however less attractive for studies concerning
the rotational evolution of stars.

\subsection{Saturation limits}
\citet{1972ApJ...171..565S} analysed the rotation and chromospheric
emission of cool stars in different evolutionary stages, and found that
their rotation rate (as well as \ion{Ca}{II} luminosity) is about
proportional to the inverse square-root of their age, that is $\Omega
\propto t^{-1/2}$.
Presuming a constant moment of inertia and a WD braking law, the
\citeauthor{1972ApJ...171..565S} relation implies a linear relationship
between the rotation rate of a star and its characteristic magnetic
field strength, $\bar{B}\propto \Omega^{n_\mathrm{de}}$ with dynamo
efficiency $n\dw{de}\simeq 1$.
However, \citet{1991LNP...380..389S} found that the magnetic flux
rather than the field strength is increasing linearly with the rotation
rate, which is now commonly adopted (at slow rotation rates).
The consequence of a continuously increasing stellar field strength is
a very efficient magnetic braking of rapidly rotating stars.
In fact, too efficient, since observations of young open clusters
reveal significant numbers of stars with rotational velocities up to $v
\sin i\sim 200\un{km/s}$ \citep{1997ApJ...479..776S}, whose existence
is, with initial rotation rates of young T Tauri stars being
observationally well constrained \citep{1993A&A...272..176B}, difficult
to explain in the framework of magnetic braking without any mechanism
which moderates the AM loss rate at higher rotation rates.
Respective studies therefore frequently presume a saturation of the AM
loss beyond a limiting rotation rate, whose value is very susceptible
to model assumptions like the internal AM redistribution; whereas some
investigations require low ($\lesssim 20\un{\Omega_\odot}$) saturation
limits \citep{1995A&A...294..469K, 1996ApJ...462..746B,
1997A&A...326.1023B, 1997ApJ...480..303K}, others argue for higher
($\gtrsim 40\un{\Omega_\odot}$) values \citep{1993ApJ...409..624S,
1994MNRAS.269.1099C}.

The saturation of the total AM loss rate is often ascribed to the
underlying dynamo mechanisms in the convective envelope, because
current dynamo theories anticipate a moderation and eventually
saturation of the amplification process ($n\dw{de}\rightarrow 0$) when
the back-reaction of strong magnetic fields suppresses plasma motions
and inhibits its further increase \citep[e.g.,][]{1993A&A...269..581R}.
However, the various theoretical models are as yet unable to provide
consistent or explicit values for the critical rotation rate or field
strength at which this occurs.

The concept of a saturation of magnetic activity is strengthened by
empirical activity-rotation relationships, which reveal (for rotation
periods longer than a few days) close correlations between the rotation
rate and the strength of activity proxies like magnetically induced
chromospheric and coronal emission \citep{1984ApJ...279..763N,
1984A&A...133..117V, 1995A&A...300..775M, 1997ApJ...479..776S,
2003A&A...397..147P}.
In rapidly rotating stars several activity signatures are found to
saturate:  the chromospheric UV emission below rotation periods $P\sim
3\un{d}$ \citep{1984A&A...133..117V}, and the EUV and (soft) X-ray
emission for $P\lesssim 2\un{d}$ \citep{1997ApJ...479..776S}.
However, the variation of photometric light curves, associated with the
presence of dark spots in the stellar photosphere, is found to increase
for even shorter rotation periods, down to $P\sim 0.35-0.5\un{d}$, for
which the chromospheric and coronal emission are already saturated
\citep{1995A&A...294..715O, 2001A&A...366..215M}; for an antithetical
point of view see \citet{1998ApJ...493..914K}.

The saturation of activity signatures is not unambiguously indicative
of a saturation of the dynamo mechanisms in the convective envelope,
since emission processes are liable to further rotation-dependent
effects like a reduction of the X-ray emitting volume through the
centrifugal stripping of hot coronal loops, or the shift of coronal
loop temperatures into different emission regimes as the effective
gravitation and pressure scale hight change with the rotation rate
\citep{1997A&A...321..177U, 1999A&A...346..883J}.
The conjecture that changes of the atmospheric emission are not
necessarily correlated with the (sub-)photospheric magnetic activity is
supported, for example, by observations of the ultra-fast rotator
VXR45a (\object{V370 Vel}, $P= 0.223\un{d})$, whose X-ray emission is
below the typical level of X-ray saturated stars
\citep{2003A&A...407L..63M}, whereas its brightness surface maps are
still very similar to those of more slowly rotating stars
\citep{2004AN....325..246M, cs13splinter}.

The large range of rotation rates ($\sim 10-70\un{\Omega_\odot}$) in
which observed activity signatures are found to saturate raises the
question whether these phenomena reflect the actual behaviour of the
underlying dynamo processes, in particular beyond which critical
rotation rate their efficiency breaks down.
Since the range of observed saturation limits practically covers the
one of suggested AM loss limits, a definite justification of the latter
in terms of a dynamo saturation is rather precarious.

\subsection{Surface magnetic flux distributions}
Doppler imaging (DI) observations of rapidly rotating stars yield
non-uniform surface brightness distributions, where, in contrast to the
case of the Sun, dark spots are not only located in equatorial regions,
but also at intermediate and polar latitudes \citep[][ and references
therein]{2002AN....323..309S}.
Theoretical models considering the formation of magnetic features at
higher latitudes involve the pre-eruptive poleward deflection of
magnetic flux inside the convection zone by the Coriolis force
\citep{1992A&A...264L..13S}, and/or its post-eruptive poleward
transport through meridional motions \citep{2001ApJ...551.1099S}.
Backed by these observational and theoretical results,
\citet{1997A&A...325.1039S} investigated the influence of a bi-modal
magnetic field distribution on the rotational evolution of cool stars.
They found that a concentration of magnetic flux at very high latitudes
reduces the total AM loss rate as efficiently as a dynamo saturation
limit at $\sim 20\un{\Omega_\odot}$.
Based on their findings, they question the concept of a dynamo
saturation at low rotation rates and argue instead for a saturation
above $50\un{\Omega_\odot}$; a similar though more qualitative argument
has also been discussed by \citet{1997ApJ...484..855B}.

\begin{figure}
\includegraphics[width=\hsize]{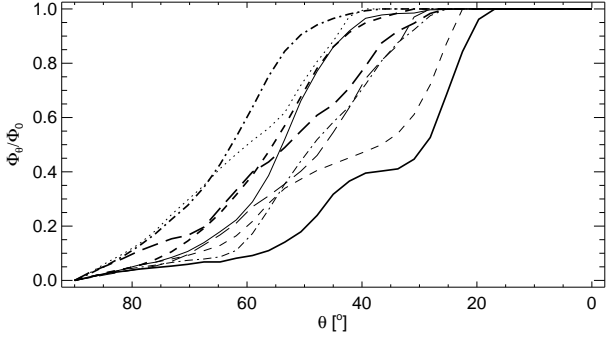} 
\caption{Cumulative open magnetic flux distributions, $\Phi_\theta / 
\Phi_0$ (cf.~Sect.~\ref{ldmsw}), on the visible hemisphere of
\object{AB Dor} ($P= 0.51\un{d}$), based on ZDI observations secured
between 1995--2003 \citep{1997MNRAS.291....1D, 1999MNRAS.302..437D,
2003MNRAS.345.1145D}.
Thin curves: 1995 ($\theta_{50\%}= 54\degr$, \emph{solid}); 1996
($47\degr$, \emph{long dashed}); 1997 ($37\degr$, \emph{short dashed});
1998 ($49\degr$, \emph{dashed dotted}); 1999 ($60\degr$,
\emph{dotted}).
Thick curves: 2000 ($29\degr$, \emph{solid}); 2001 ($53\degr$,
\emph{long dashed}); 2002 ($56\degr$, \emph{short dashed}); 2003
($63\degr$, \emph{dashed dotted}).}
\label{cuflobs.fig}
\end{figure}

The work of \citeauthor{1997A&A...325.1039S} was focused on flux
concentrations around the pole, assuming that the observed brightness
distributions indicate likely locations of open field lines along which
a stellar wind can escape.
But the bare existence of starspots does not a priori imply information
about the associated magnetic field topology, that is neither about the
existence nor about the amount of open magnetic flux.
Zeeman-Doppler imaging (ZDI) observations in contrast directly confirm
the magnetic origin of the dark features and enable a determination of
the magnetic field at the stellar surface \citep{1997MNRAS.291..658D}.
Such field distributions serve as boundary conditions for field
extrapolation techniques \citep{1969SoPh....9..131A}, which reveal
large-scale magnetic field topologies and consequently the distribution
of closed and open magnetic field lines
\citep[e.g.,][]{2002MNRAS.333..339J, 2002MNRAS.336.1364J}.
In the case of the rapidly rotating star \object{LQ Hya} ($P=
1.6\un{d}$) the latitudinal distributions of open flux show that large
amounts are located at intermediate and high latitudes
\citep{2004MNRAS.355.1066M}.
A similar result is found in the case of \object{AB Dor} ($P=
0.51\un{d}$) for observations between 1995 and 2003
\citep{1997MNRAS.291....1D, 1999MNRAS.302..437D, 2003MNRAS.345.1145D}:
Figure \ref{cuflobs.fig} shows the (normalised) cumulated open magnetic
flux, integrated from the equator to a co-latitude $\theta$.
The curves indicate that $50\%$ of the total open magnetic flux of a
hemisphere is on average located below/above $\sim 45\degr$, with
variations of $\pm 15\degr$ depending on the observational epoch.
These results show that the distribution of open magnetic flux can be
distinctively different from spot distributions determined from surface
brightness maps alone.

The present work extends the study of \citet{1997A&A...325.1039S} and
investigates the impact of latitude-dependent flux distributions on the
rotational evolution of solar-type stars in more detail.
Using the magnetic wind model described in \citet{2005smaamldtlmw}, we
quantify the influence of prescribed flux distributions with different
degrees of non-uniformity on the rotational evolution of stars, and
verify their importance by comparing their impact with the influence of
other magnetic field-related model parameters.

\section{Model setup}
\label{mose}
The stellar structure of the low-mass star considered here consists of
an outer convective envelope and an inner radiative core, each taken to
be in solid-body rotation with possibly different rotation rates.
With $J= I \Omega$ being the AM, $I$ the moment of inertia, and
$\Omega$ the rotation rate, the rotational evolution of the star is
determined by the set of coupled differential equations
\begin{eqnarray}
\frac{d \lg \Omega\dw{e}}{d \lg t}
& = &
-
\frac{d \lg I\dw{e}}{d \lg t}
+
\frac{t}{J\dw{e}} 
\left( \dot{J}\dw{e,\dot{M}} + \dot{J}\dw{e,RI} + \dot{J}\dw{W} \right)
\label{lgome}
\\
\frac{d \lg \Omega\dw{c}}{d \lg t}
& = &
- 
\frac{d \lg I\dw{c}}{d \lg t}
+
\frac{t}{J\dw{c}} 
\left( \dot{J}\dw{c,\dot{M}} + \dot{J}\dw{c,RI} \right)
\ ,
\label{lgomi}
\end{eqnarray}
which comprise changing moments of inertia and the AM transfer across 
appropriate boundaries; indices `e' and `c' denote quantities of
the envelope and of the core, respectively.

\subsection{Internal angular momentum redistribution}
\label{iamr}
\begin{figure}
\includegraphics[width=\hsize]{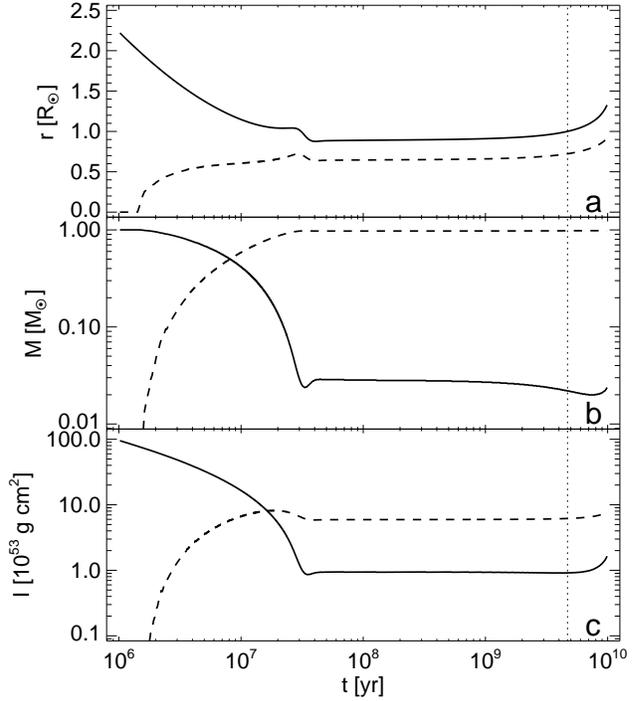}
\caption{Evolution of the outer radii, $r\dw{e/c}$ (Panel {\bf a}),
masses, $M\dw{e/c}$ (Panel {\bf b}), and moments of inertia,
$I\dw{e/c}$ (Panel {\bf c}) of the convective envelope (\emph{solid})
and radiative core (\emph{dashed}) of a $1\un{M_{\sun}}$ star.
For $t\lesssim 1.5\un{Myr}$ the star is fully convective.
Its MS phase starts at $t\simeq 40\un{Myr}$; the vertical dashed line 
marks the age of the Sun.}
\label{stepar.fig}
\end{figure}

The stellar structure is taken to be spherically symmetric.
The temporal change of the moments of inertia is determined by an
evolutionary sequence of stellar models of a $1\un{M_{\sun}}$ star,
which was generated with an updated version of the stellar evolution
code of \citet{1967micp..7..129}.
Figure \ref{stepar.fig} shows the evolution of the (outer) radii, 
masses, and moments of inertia of both the convective envelope and the
radiative core.

As the core-envelope interface recedes outward, the dynamical stability
properties at the base of the convection zone change and originally
unstable mass settles down on the radiative core.
This mass transfer is accompanied by an AM transfer,
\begin{equation}
\dot{J}\dw{e,\dot{M}}
= 
- \frac{2}{3} r\dw{c}^2 \Omega\dw{e} \dot{M}\dw{c}
= 
- \dot{J}\dw{c,\dot{M}}
\ .
\label{jdot_mdot}
\end{equation}
The temporal change of the core mass, $\dot{M}\dw{c}$, is determined 
from the sequence of stellar models (Fig.\ \ref{stepar.fig}b); since
the mass loss of the envelope through the stellar wind is negligibly
small, it is $\dot{M}\dw{e}= - \dot{M}\dw{c}$.

Hydrodynamic and hydromagnetic interaction at the boundary between the 
radiative and convective regions are expected to entail a coupling 
between the core and the envelope.
Different coupling mechanisms based on magneto-viscous interaction or
large-scale internal circulations have been investigated by
\citet{1989ApJ...338..424P, 1993ApJ...417..762C, 1997A&A...326.1023B,
1998A&A...333..629A} and references therein.
Here, a more qualitative core-envelope coupling model is used, following
the parametric approach of \citet{1991ApJ...376..204M} and
\citet{1995A&A...294..469K}.
Based on the Rayleigh criterion a dynamical stable state of rotation
inside a star requires\footnote{in the equatorial plane and in the
absence of viscosity} the increase of the specific AM with increasing 
radius, $d (\Omega r^2) / dr> 0$.
Given that the rotation rate of the convective envelope is braked and
smaller than the rotation rate of the core, the stability condition is
violated at the core-envelope interface and a rotational instability
sets in, transferring AM,
\begin{equation}
\dot{J}\dw{e,RI}
=
\frac{\Delta J}{\tau\dw{c}}
= 
- \dot{J}\dw{c,RI}
\ ,
\label{jdot_ri}
\end{equation}
from the core to the envelope to eliminate the differential rotation
between the two regions.
The AM required to equalise the two rotation rates,
\begin{equation}
\Delta J
=
\frac{I\dw{e} J\dw{c} - I\dw{c} J\dw{e}}{I\dw{e} + I\dw{c}}
=
\frac{I\dw{e} I\dw{c}}{I\dw{e} + I\dw{c}} 
\left( \Omega\dw{c} - \Omega\dw{e} \right)
\ ,
\label{deltaj}
\end{equation}
is transferred on a timescale, $\tau\dw{c}$, which is supposed to 
characterise the various visco-magnetic coupling mechanisms.
The coupling time quantises the possibility to deposit AM in the core 
during the PMS phase and its retarded transfer to the envelope in the
course of the MS evolution.
For the sake of simplicity the value of $\tau\dw{c}$ is taken to be 
constant during the entire rotational evolution of the star.

\subsection{Latitude-dependent magnetised stellar winds}
\label{ldmsw}
The AM loss rate of the convective envelope through a
latitude-dependent magnetised wind is determined following the approach
of \citet{2005smaamldtlmw}.
The stationary, polytropic stellar wind is assumed to be symmetric with
regard to both the rotation axis and the equator.
The poloidal component of the magnetic field is taken to be radial, 
with field lines forming spirals around the rotation axis on coni with
constant opening angles.
The whole stellar surface contributes to the wind, without `dead zones'
retaining mass from escaping \citep[cf.][]{1987MNRAS.226...57M}.

\begin{figure}
\includegraphics[width=\hsize]{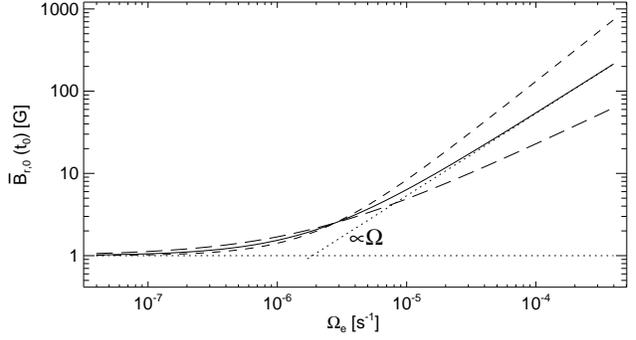}
\caption{Dependence of the surface averaged radial magnetic field 
strength, $\bar{B}_{r,0} (t_0)$, on the rotation rate, $\Omega\dw{e}$, 
of the convective envelope.
For large rotation rates the dependencies follow approximately linear
(\emph{solid}), sub-linear (\emph{long dashed}), or super-linear 
(\emph{short dashed}) power laws, $\propto \left( \Omega\dw{e} /
\Omega\dw{e,\odot} \right)^{n_\Omega}$, with $n_\Omega= 1, 0.75$, and 
$1.25$, respectively.
The deviation from the power law at small rotation rates is due to the
constant background field strength.}
\label{omemag.fig}
\end{figure}

\begin{figure}
\includegraphics[width=\hsize]{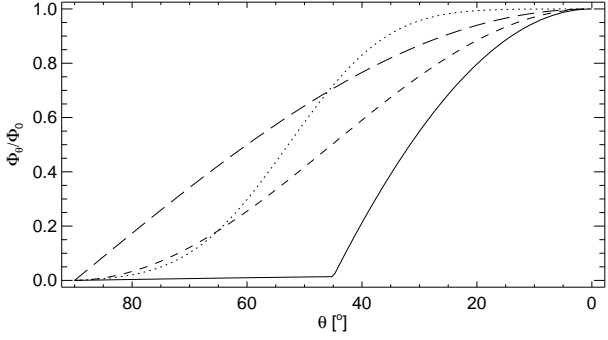}
\caption{Cumulative open magnetic flux, $\Phi_\theta/\Phi_0$, of the 
Coronal Hole (\emph{solid}, for $\theta_0= 45\degr$) and Latitudinal 
Belt (\emph{dotted}, for $\theta_0= 50\degr$) model.
Respective curves for a constant (\emph{long dashed}) and dipolar
(\emph{short dashed}) field distribution are show for comparison; see
also Fig.\ \ref{cuflobs.fig} for observational results.}
\label{cumuflux2.fig}
\end{figure}

The magnetic wind properties are determined through the radial magnetic
field strength,
\begin{equation}
B_{r,0} \left( t, \theta, \Omega_{\rm e} \right)
=
\left( \frac{r_0 \left( t_0 \right)}{r_0 \left( t \right)} \right)^2
\big(
 B_< 
 + 
 \Delta B \left( \Omega_{\rm e} \right) \cdot f \left( \theta \right)
\big)
\ ,
\label{defbr0}
\end{equation}
given at a reference level, $r_0$, close to the stellar surface.
The time-dependent radius ratio is to ensure that the total magnetic 
flux,
\begin{equation}
\Phi_0 \left( \Omega\dw{e} \right)
=
\int\limits_0^{2\pi}
\int\limits_0^\pi B_{r,0} r_0^2 \sin \theta\, d \theta\, d \phi
=
4 \pi r_0^2 (t_0) \bar{B}_{r,0} \left( t_0, \Omega\dw{e} \right)
\ ,
\label{defphi0}
\end{equation}
only depends on the rotation rate of the star 
\citep{1991LNP...380..389S}, and an arbitrary reference time, $t_0$.
The efficiency of the underlying dynamo mechanism is expected to 
increase with the rotation rate of the convection zone.
The field strength variation, $\Delta B$, is therefore determined in 
a way that the \emph{surface averaged} field strength, 
\begin{equation}
\bar{B}_{r,0} \left( t_0, \Omega\dw{e} \right) 
= 
\int\limits_0^{\pi/2} B_{r,0} \sin \theta\, d\theta
\ ,
\label{defbr0avg}
\end{equation}
obeys the functional behaviour shown in Fig.\ \ref{omemag.fig}.
The non-uniform flux distributions, superposed on a constant 
`background' field, $B_<$, are characterised by an enhancement of 
magnetic flux at non-equatorial latitudes (Fig.\ \ref{bmodels.fig}):
\begin{figure}
\includegraphics[width=\hsize]{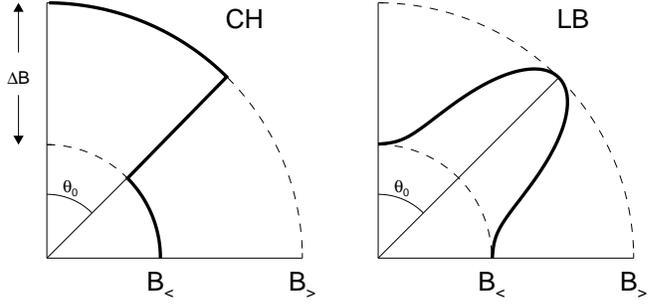}
\caption{Latitude-dependent magnetic field distributions, $f(\theta)$, 
close to the stellar surface.
The non-uniformity of the Coronal Hole (\emph{CH}) and Latitudinal Belt
(\emph{LB}) model is parametrised through the co-latitude $\theta_0$.
\emph{Dashed lines} show the lower, $B_<$, and upper, $B_>= B_< + 
\Delta B$, field strengths of each field distribution.}
\label{bmodels.fig}
\end{figure}
\begin{itemize}
\item
Coronal Hole (CH) model
\begin{equation}
f (\theta)
=
\left\{ \begin{array}{cl}
1 & \textrm{for $0\le \theta\le \theta_0$} \\
0 & \textrm{for $\theta_0< \theta< 90\degr$}
\end{array} \right.
\label{defbch}
\end{equation}
\item
Latitudinal Belt (LB) model
\begin{equation}
f (\theta) = \cos^{16} \left( \theta - \theta_0 \right)
\label{defblb}
\end{equation}
\end{itemize}
Whereas the Latitudinal Belt model closely resembles what is found from
field extrapolations based on ZDI images \citep{2004MNRAS.355.1066M},
the Coronal Hole model is motivated by numerous DI surface brightness
maps \citep[e.g.,][]{2002AN....323..309S}.
Different degrees of high-latitude flux concentrations are realised by 
changing the co-latitude $\theta_0$ of the analytically prescribed
functions $f$ (Fig.\ \ref{cumuflux2.fig}).
We classify non-uniform flux distributions by the location of their 
$50\%$-open flux level, $\theta_{50\%}$, where the cumulated open
magnetic flux, $\Phi_\theta= 4 \pi \int_{\pi/2}^\theta B_{r,0} r_0^2
\sin \theta' d \theta'$, reaches half of the total value, $\Phi_0$.

The total, surface integrated AM loss rate,
\begin{eqnarray}
\dot{J}\dw{W}
& = &
\dot{J}_{\rm WD} \frac{3}{2} \int \limits_0^{\pi/2}
\left( \frac{r_{\rm A}}{\bar{r}_{\rm A}} \right)^4 
\left( \frac{\rho_{\rm A}}{\bar{\rho}_{\rm A}} \right)
\left( \frac{v_{r,{\rm A}}}{\bar{v}_{r,{\rm A}}} \right)
\sin^3 \theta d \theta
\ ,
\label{taml}
\end{eqnarray}
is expressed in terms of the plasma density, $\rho_{\rm A}$, and radial
flow velocity, $v_{r,{\rm A}}$, at the Alfv\'enic point, $r_{\rm A}$,
where the flow velocity of the wind equals the Alfv\'en velocity
\citep[cf.][]{2005smaamldtlmw}.
Whereas $\rho_{\rm A}, v_{r,{\rm A}}$, and $r_{\rm A}$ are functions of
the co-latitude, $\theta$, and subject to the latitude-dependent field
distributions, Eq.\ (\ref{defbr0}), the respective quantities,
$\bar{\rho}_{\rm A}, \bar{v}_{r,{\rm A}}$, and $\bar{r}_{\rm A}$, are
determined in the equatorial plane ($\theta= \pi/2$) using the surface
averaged field strength, Eq.\ (\ref{defbr0avg}).
The latter quantities determine the AM loss rate following the approach
of \citet{1967ApJ...148..217W},
\begin{equation}
\dot{J}_{\rm WD} = \frac{8\pi}{3} \Omega \bar{r}_{\rm A}^4 
\bar{\rho}_{\rm A} \bar{v}_{r,{\rm A}}
\ ,
\label{tamlwd}
\end{equation}
which is based on the simplifying assumption that the equatorial wind 
structure can be generalised to all latitudes.

\subsection{Reference model parameters}
\label{rmp}
The wind structure is determined through boundary conditions prescribed
at the reference level $r_0(t)= r\dw{e} (t) + \Delta r$, over the range
of co-latitudes $0< \theta\le \pi/2$, measured from the stellar north
pole.
$r\dw{e}(\equiv R_\ast)$ is the time-dependent outer radius of the
convective envelope (Fig.\ \ref{stepar.fig}a), and $\Delta r=
0.1\un{R_{\sun}}$ a constant radial offset to locate the reference
level of the wind at the base of the corona.

For cool stars other than the Sun thermal wind properties are poorly
constrained by observations; for possible constraints resulting from
relationships between the temperature and density of closed coronal
loops and rotation/Rossby number see, for example,
\citet{1991MNRAS.252P..21J, 2003ApJ...599..516I}.
In the following solar-like values are assumed for the temperature,
$T_0= 2\cdot10^6\un{K}$, and (particle) density, $n_0=
10^8\un{cm^{-3}}$.
The entropy change of the wind with increasing distance from the star
is quantified through the polytropic index, $\Gamma= 1.15$.

The surface averaged magnetic field strength, defined in
Eq.\ (\ref{defbr0avg}), is taken to follow the quasi-linear power law
shown in Fig.\ \ref{omemag.fig} (solid line), that is $\bar{B}_{r,0}=
\left( 1 + 1.5 \left( \Omega\dw{e} / \Omega_\odot \right) \right)
\un{G}$, with $\Omega_\odot= 2.8\cdot10^{-6}\un{s^{-1}}$.
This is in agreement with observations of rapidly rotating stars, which
show magnetic field strengths up to about two orders of magnitude
larger than in the case of the Sun \citep{1997MNRAS.291....1D}.

Helioseismological observations show that the rotation rate in the 
solar interior is roughly uniform \citep{2003ARA&A..41..599T}.
The rotation rate of the solar radiative core and convection zone are
within $\sim 4\%$ about $\Omega_{\odot}= 2.8\cdot10^{-6}\un{s\up{-1}}$.
The present Sun thus constrains the coupling timescale to the effect
that the value of $\tau\dw{c}$ ought to achieve isorotation within a
few percent at the solar age, $t_{\sun}\approx 4.7\un{Gyr}$.
Using the reference model parameters described above, we accomplished
simulations with different coupling timescales to determine the
relative deviation, $(\Omega\dw{c} - \Omega\dw{e}) / \Omega\dw{c}$,
from isorotation at solar age (Table \ref{maxerr.tbl}).
\begin{table}
\caption{Mean deviation, $\left( \Omega\dw{c} - \Omega\dw{e} \right) / 
\Omega\dw{c}$, from isorotation at solar age, $t_{\odot}= 4.7\un{Gyr}$,
for different coupling timescales, $\tau\dw{c}$.}
\label{maxerr.tbl}
\begin{tabular}{rccccccc}
\hline
$\tau\dw{c}$ [Myr] & 1 & 5 & 10 & 15 & 20 & 50 & 100 \\
$\Delta \Omega / \Omega\dw{c}$ [\%] & 0.3 & 1.3 & 2.6 & 4.1 & 5.6 & 
16.4 & 35.7 \\
\hline
\end{tabular}
\end{table}
For $\tau\dw{c}\simeq 15\un{Myr}$, a value similar to the one adopted
by \citet{1991ApJ...376..204M} or \citet{1995A&A...294..469K}, the
deviation is in accord with the observational constrains given by the
Sun.

\section{Results}
\label{resu}
We determine the rotational evolution of $1\un{M_{\sun}}$ stars with 
initial rotation rates $\Omega_0= 5\cdot10^{-6}$, $2\cdot10^{-5}$, and 
$6\cdot10^{-5}\un{s^{-1}}$ from the age $1.6\un{Myr}$ onward.
The rotation periods, between $1.2\un{d}$ and $15\un{d}$, approximately
span the observed range of rotation periods of young $(\lesssim
5\un{Myr})$ T Tauri stars \citep{1993A&A...272..176B}.
In their initial state the convective and radiative zones are taken to 
be in isorotation, that is $\Omega\dw{c}= \Omega\dw{e}= \Omega_0$.

\begin{figure}
\includegraphics[width=\hsize]{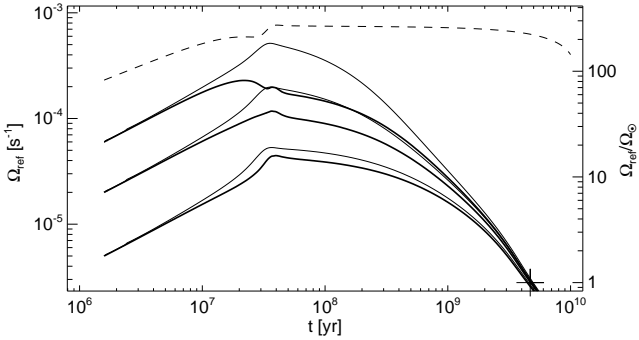} 
\caption{Rotational evolution of the convective (\emph{thick}) and
radiative (\emph{thin}) zone of stars subject to an AM loss rate
according to \citet{1967ApJ...148..217W}.
The \emph{dashed} line indicates the break-up rotation rate, for which
the co-rotation radius equals the actual stellar radius,
$\Omega\dw{bu}= \sqrt{ G M_\ast / r_\ast^3 }$.
The \emph{cross} marks the rotation rate of the present Sun.
}
\label{reference.fig}
\end{figure}

The rotational histories determined with the AM loss rate
$\dot{J}\dw{W}= \dot{J}\dw{WD}$, that is using the WD approach,
Eq.\ (\ref{tamlwd}), are used as reference cases
(Fig.\ \ref{reference.fig}).
In the course of the PMS evolution the rotation rate of the envelope is
dominated by its decreasing moment of inertia and the AM loss which
goes with the mass settling down onto the radiative core, whereas
during the MS phase only the magnetic braking and the internal coupling
are relevant.

\subsection{Non-uniform flux distributions}
We determine the rotational history of stars subject to the Latitudinal
Belt (LB) and Coronal Hole (CH) flux distributions with $50\%$-open
flux levels located at different latitudes.
At first, the non-uniform flux patterns are taken to be stationary
($\theta_{50\%}= \mathrm{const.}$) during the entire evolution, to
separate their influence from other rotation-dependent effects.

\begin{figure*}
\includegraphics[width=0.49\hsize]{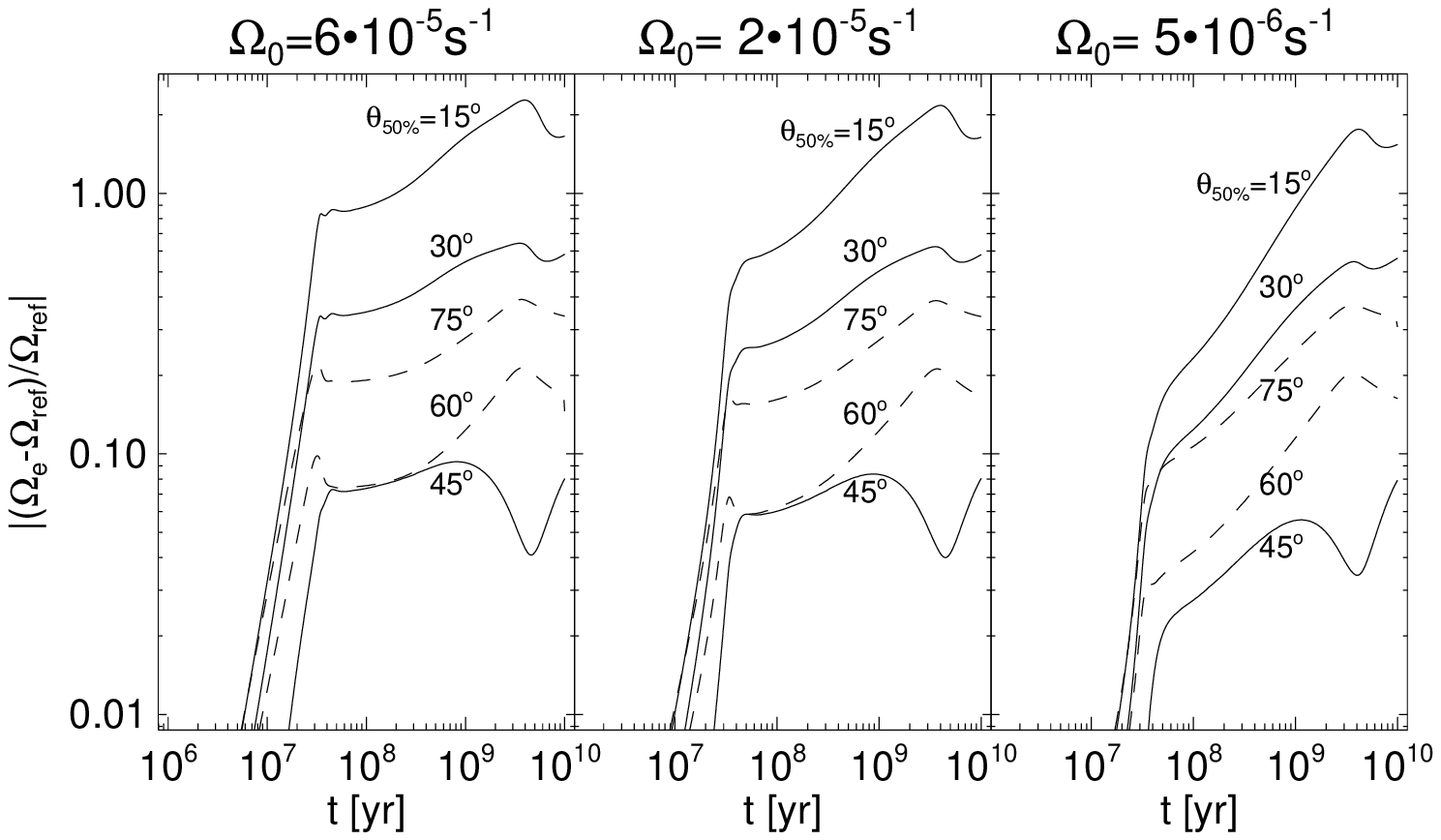} \hfill
\includegraphics[width=0.49\hsize]{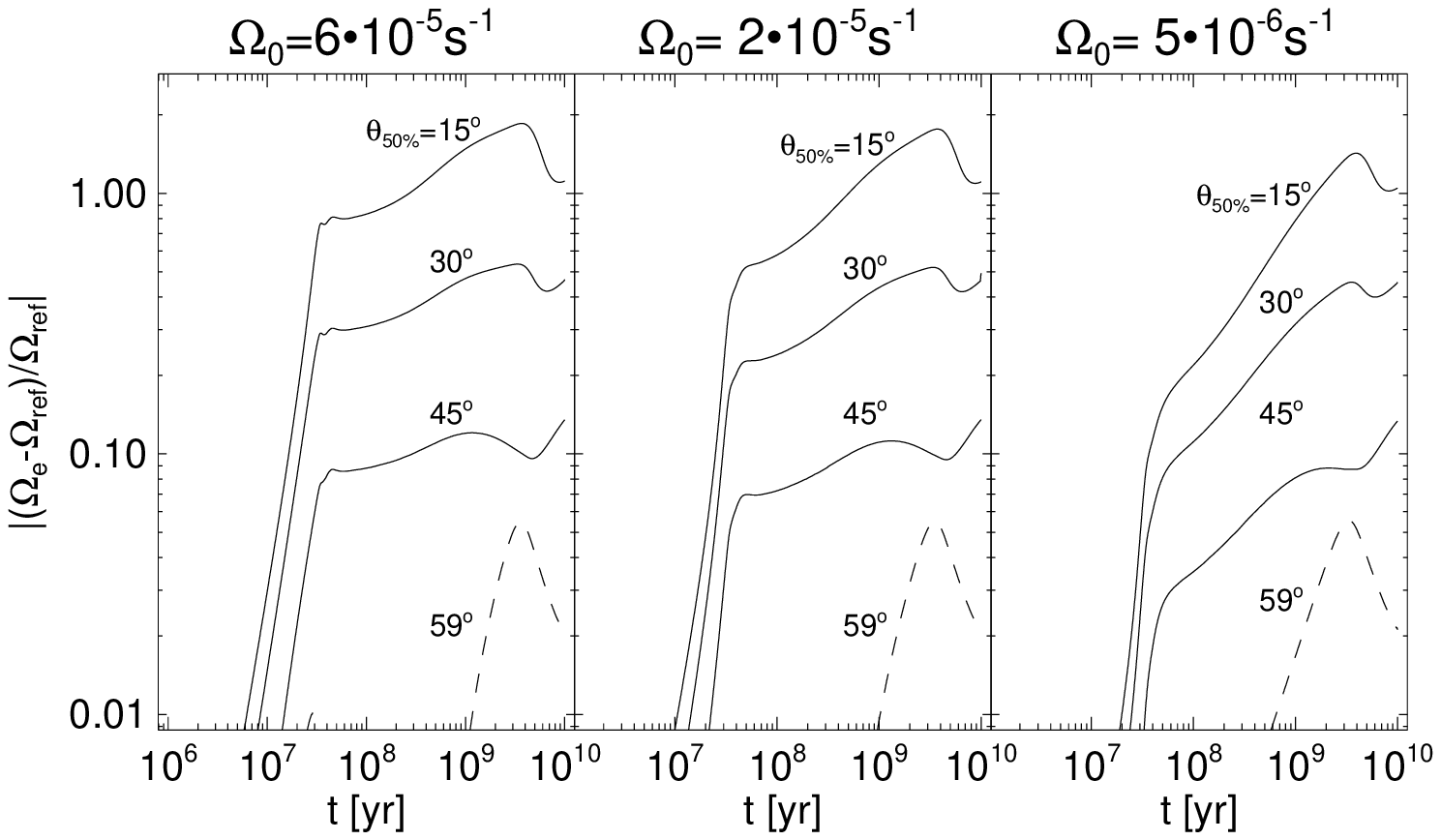} 
\caption{Relative deviations of the envelope rotation rate for the
Latitudinal Belt (\emph{left}) and Coronal Hole (\emph{right}) flux
distributions.
For $50\%$-open flux levels at low latitudes (i.e., large co-latitudes
$\theta_{50\%}$, \emph{labels}) the deviations are negative
(\emph{dashed}), and for high latitudes positive (\emph{solid}).}
\label{scanC2&6.fig}
\end{figure*}

\begin{figure}
\includegraphics[width=\hsize]{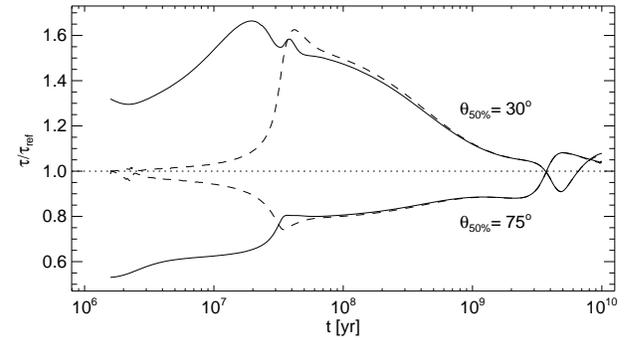} 
\caption{Relative timescales (with respect to respective reference
cases) on which the rotation rate of the envelope changes due to the
magnetic braking (\emph{solid}) and internal coupling (\emph{dashed}),
for $\Omega_0= 2\cdot10^{-5}\un{s^{-1}}$ and a LB distribution with
$\theta_{50\%}= 30$ and $75\degr$.}
\label{tauras.fig}
\end{figure}

An accumulation of open magnetic flux at high (low) latitudes causes,
with respect to an uniform flux distribution, a reduction (enhancement)
of the AM loss rate through the stellar wind.
At the age of the present Sun the resulting deviations of the stellar
rotation rate are found to cover a range of values between about
$-40\%$ and $200\%$ different than the respective reference cases
(Fig.\ \ref{scanC2&6.fig}).
The relative deviations show two distinctive regimes, corresponding to 
the PMS and the MS phase.
For the deviations to increase (decrease) with time the net AM loss
rate of the envelope has to be smaller (larger) than in the reference
case.
Since we consider the difference between two rotational histories, the
rotation-independent mechanisms cancel out, so that only the magnetic
braking and the internal coupling are relevant.
The magnetic braking is a direct consequence of the present stellar 
rotation and causes an immediate AM loss of the envelope.
The internal coupling, in contrast, is proportional to the differential
rotation and during the early evolution rather inefficient, due to the
initial isorotation (Fig.\ \ref{tauras.fig}).
In the PMS phase the difference between the AM loss of the envelope 
and its internal AM gain is consequently large and the deviations of 
the rotation rate quickly increasing.
After arrival on the MS the differential rotation is adjusted to the
new situation and the tapping of the AM reservoir of the more rapidly
rotating core now replenishes most of the AM loss carried away by the
magnetised wind; the further increase of the deviations is respectively
weaker.

The actual spin-down time scale of the envelope is longer than the
braking time scale, because the AM gain of the envelope through the
internal coupling follows closely its AM loss through the magnetic
braking (Fig.\ \ref{tauras.fig}).
Note that in the course of the evolution the latter is found to 
converge toward the reference value.
Since the magnetic field strength decreases with the rotation rate the
relative contribution of the (latitude-dependent) magneto-centrifugal
driving to the overall wind acceleration becomes smaller and the one of
the (latitude-independent) thermal driving larger.
The difference between the \citeauthor{1967ApJ...148..217W} and the
present non-uniform wind approach thus becomes smaller and the
spin-down timescale similar.

The non-uniform flux distributions with $\theta_{50\%}= 60\degr$ have
nominally the same $50\%$-flux level as the uniform flux distribution
of the reference cases.
The deviations of the relative rotation rate of $\sim 10\%$ in the case
of the LB model indicate that the usage of $\theta_{50\%}$ only allows
for a rudimentary classification of flux patterns.
By comparing the rotational histories resulting from different flux
distributions with equivalent $50\%$-open flux levels, we find that
over the range $\theta_{50\%}= 10-60\degr$ the peaked Latitudinal Belt
and the bi-modal Coronal Hole model yield rotational evolutions which
are within a tolerance of $15\%$ consistent.
In the special case of a dipolar field distribution (with $f (\theta)=
\cos \theta$ and $\theta_{50\%}= 45\degr$, Fig.\ \ref{cumuflux2.fig}),
the correspondency is found to be good within about $10\%$.

\begin{figure}
\includegraphics[width=\hsize]{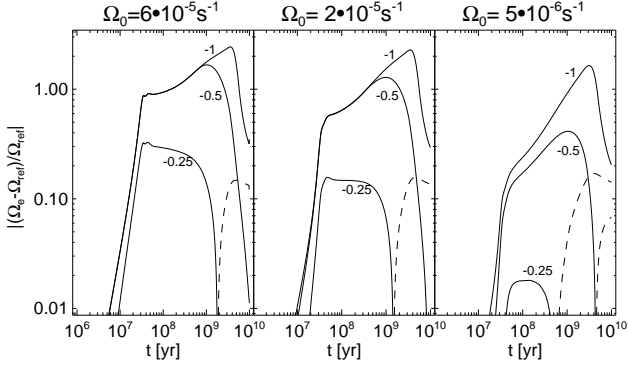} 
\caption{Relative deviations (\emph{dashed} lines indicate negative 
values) in the case of rotation-dependent $50\%$-open flux levels 
$\theta_{50\%}= 90\degr \left( \Omega / \Omega_\odot 
\right)^{n_\mathrm{rd}}$, with $n\dw{rd}= -0.25, -0.5$, and $-1$
(\emph{labels}).
Its value is constrained to the range $15\degr\le \theta_{50\%}\le
60\degr$.}
\label{tideC2.fig}
\end{figure}

Observations show that surface distributions of starspots depend on the
stellar rotation rate, with spots being preferentially located at 
higher latitudes the faster the star rotates.
We investigate this aspect by assuming a rotation-dependent latitude of
the $50\%$-flux level, which follows the power law $\theta_{50\%}=
90\degr \left( \Omega / \Omega_\odot \right)^{n_\mathrm{rd}}$, with
$n\dw{rd}= -0.25, -0.5$, and $-1$; this simple relationship is only
used to examine the basic effects and not meant to reproduce any
complex observed or theoretically derived relations.
To avoid unrealistic high flux concentrations in the vicinity of the
poles or the equator, the $50\%$-flux latitudes are constrained to the
range $15\degr\le \theta_{50\%}\le 60\degr$
\citep[cf.][]{1997A&A...325.1039S}.
Figure \ref{tideC2.fig} shows the relative deviations of the rotation
rates from the reference values in the case of a rotation-dependent LB 
distribution; for the CH model the results are very similar.
The major consequence of the rotation-dependence is the enhanced 
stellar spin-up and spin-down during the PMS and the late MS phase, 
respectively.
The evolutionary stage at which the rotational history is altered by
the shifting of magnetic flux to lower latitudes depends
however on the actual functional dependence of $\theta_{50\%}
(\Omega)$.

\subsection{Comparison with dynamo efficiency and saturation}
To examine whether the impact of non-uniform flux distributions is
significant with respect to other magnetic field-related model
assumptions, we compare the deviations described above with those
obtained by successively changing the functional dependence of the
magnetic field strength on the stellar rotation rate in the framework
of the reference cases (i.e., with uniform flux distributions).

\begin{figure}
\includegraphics[width=\hsize]{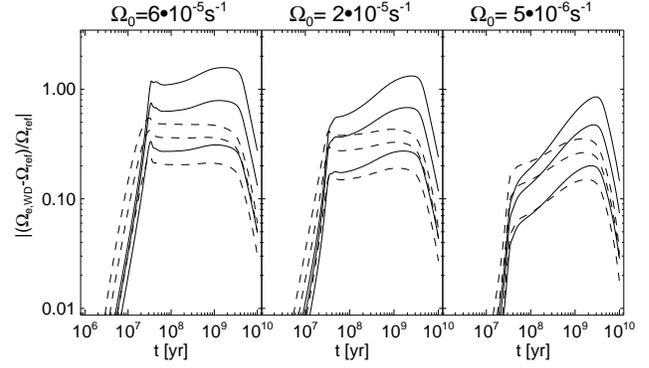} 
\caption{Relative deviations in the case of non-linear dynamo 
efficiencies, $\Delta B\propto \Omega^{n_{\rm de}}$.
For $n\dw{de}= 0.7, 0.8$, and $0.9$ (\emph{solid, top down}) the
deviations are positive, and for $n\dw{de}= 1.3, 1.2$, and $1.1$
(\emph{dashed, top down}) negative.}
\label{dyef2.fig}
\end{figure}

\begin{figure}
\includegraphics[width=\hsize]{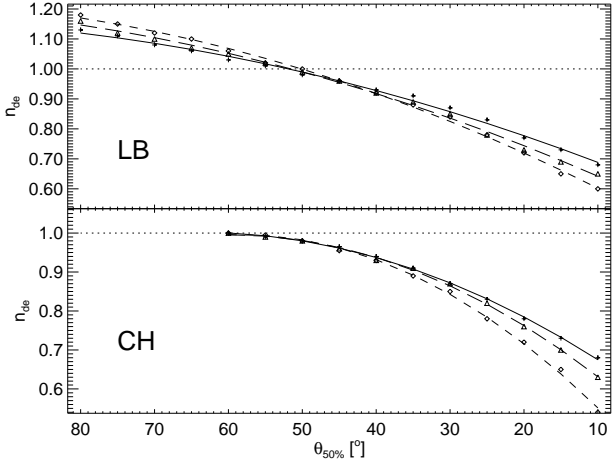} 
\caption{Relationships between the dynamo efficiency, $n\dw{de}$, and
the latitude, $\theta_{50\%}$, of the $50\%$-open flux level in the
case of the Latitudinal Belt (LB, \emph{top}) and Coronal Hole (CH,
\emph{bottom}) distributions.
The initial stellar rotation rates are $\Omega_0= 6\cdot10^{-5}$
(\emph{solid, crosses}), $2\cdot10^{-5}$ (\emph{long dashed,
triangles}), and $5\cdot10^{-6}\un{s^{-1}}$ (\emph{short dashed,
rhombs}).}
\label{scdycmp.fig}
\end{figure}

A linear dynamo efficiency is an acceptable approximation in the case
of slowly rotating stars, but it is found to fail for fast rotators.
We analyse the sensitivity of the rotational evolution on changes of
the dynamo efficiency by re-calculating the reference cases, but now
subject to non-linear magnetic field-rotation relations, $\Delta
B\propto \Omega\dw{e}^{n_{\rm de}}$, with $n\dw{de}\neq 1$
(cf.\ Fig.\ \ref{omemag.fig}).
During the MS phase, the resulting relative deviations cover a large 
range between $-50\%$ and $150\%$ (Fig.\ \ref{dyef2.fig}).
Sub-linear (super-linear) dynamo efficiencies imply a weaker (stronger)
increase of the magnetic field strength with rotation rate and
consequently a moderation (enhancement) of the magnetic braking.
Their influence is therefore qualitatively similar to
rotation-dependent concentrations of magnetic flux at high and low
latitude, respectively.
To determine explicit relations between the dynamo efficiency and the
location of the $50\%$-open flux level, we generate grids of rotational
histories with different $n\dw{de}$- and $\theta_{50\%}$-values, and
associate corresponding values by comparing the relative derivations of
the rotation rate from the reference case\footnote{Since the variations
of the rotational histories over the entire evolutionary time range are
qualitatively different, we constrain the comparison to the early MS
phase between $50\un{Myr}\le t\le 800\un{Myr}$.}.
Figure \ref{scdycmp.fig} shows quadratic fits of the resulting
grid-based relations for the LB and CH flux distribution.
The relationships show that non-uniform flux distributions can imitate
a large range (here: between $-45\%$ and $15\%$) of dynamo efficiencies
and, consequently, if not taken properly into account, conceal them
from an observational determination.

\begin{figure}
\includegraphics[width=\hsize]{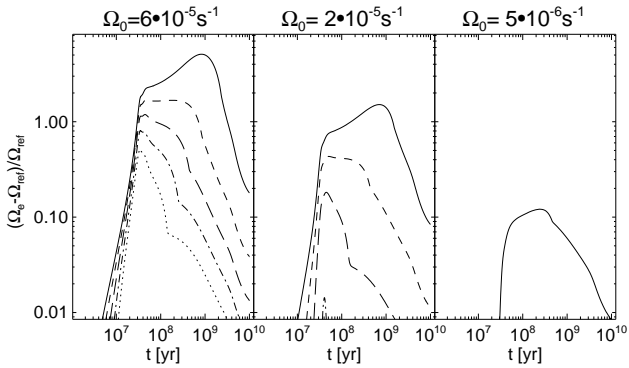} 
\caption{Relative deviations of the envelope rotation rate in the case
of a dynamo saturation beyond $\Omega\dw{sat}= 10$ (\emph{solid}), 
$20$ (\emph{long dashed}), $30$ (\emph{short dashed}), $40$ 
(\emph{dashed-dotted}), and $50\un{\Omega_\odot}$ (\emph{dotted}).
}
\label{dysa3.fig}
\end{figure}

\begin{figure}
\includegraphics[width=\hsize]{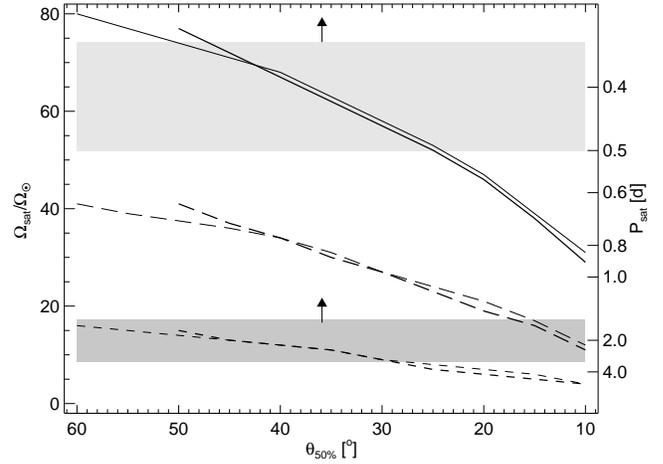} 
\caption{Relationships between the nominal saturation rotation rate,
$\Omega\dw{sat}$, and the latitude, $\theta_{50\%}$, of the $50\%$-open
flux level for stars with initial rotation rates $\Omega_0=
6\cdot10^{-5}$ (\emph{solid}), $2\cdot10^{-5}$ (\emph{long dashed}),
and $5\cdot10^{-6}\un{s^{-1}}$.
In contrast to the Coronal Hole model (\emph{thin lines}), Latitudinal
Belt distributions (\emph{thick lines}) with $\theta_{50\%}\gtrsim
53\degr$ have no corresponding dynamo saturation.
The shaded stripes indicate the rotational regimes where the
chromospheric and coronal emission (\emph{dark grey}) and the variation
of the photometric light curve (\emph{light grey}) are observationally
found to saturate.
}
\label{scdscmp.fig}
\end{figure}

The limiting field strength beyond which dynamo saturation occurs is
referred to in terms of a critical rotation rate, $\Omega\dw{sat}$,
that is $\Delta B (\Omega\dw{e} > \Omega\dw{sat})= \Delta B
(\Omega\dw{sat})$.
We repeat our calculations of the reference cases under the assumption
of saturation rotation rates in the range of $\Omega\dw{sat}=
2-80\un{\Omega_\odot}$ (see Fig.\ \ref{dysa3.fig} for examples).
The saturated magnetic braking causes an enhanced spin-up during the
PMS phase, higher rotation rates on the ZAMS, and a weaker spin-down
during the MS phase until the rotation rate descends into the regime of
the non-saturated dynamo.
We compare the rotational histories resulting from a saturated magnetic
braking with those resulting from non-uniform flux distributions to
associate $50\%$-open flux level with corresponding dynamo saturation
rates (Fig.\ \ref{scdscmp.fig}).
For given initial stellar rotation rates and flux distributions, the
curves\footnote{Note that this comparison considers exclusive
mechanisms only, that is how a saturated dynamo with a uniform flux
distribution corresponds to an unconstrained dynamo with non-uniform
flux distribution.} describe \emph{nominal} dynamo saturations which
cause similar rotational histories like a non-uniform flux distribution
with a $50\%$-open flux level at the respective co-latitude.
If the \emph{actual} dynamo saturation rate is larger than then nominal
one, the rotational evolution is likely to be dominated by the impact
of the non-uniform flux distribution.
If, in contrast, the actual dynamo saturation rate is smaller than the
nominal one, then the behaviour of the magnetic dynamo is expected to
dominate the evolution.
A concentration of magnetic flux at high latitudes of rapidly rotating
stars can therefore imitate a rotational evolution like a dynamo
saturation limit of about $30-40\un{\Omega_\odot}$.
In case a considerable amount of open flux is located at intermediate
or low latitudes, as indicated by recent field reconstructions based on
ZDI observations, the nominal saturation level would be rather high
($\gtrsim 60\un{\Omega_\odot}$) and therefore probably beyond the
actual dynamo saturation limit.
In this case non-uniform flux distributions alone could not account for
the very high rotation rates of young stars.

\section{Discussion}
\label{disc}
Our results show that a concentration of open magnetic flux at high
latitudes of active stars yields rotational histories which deviate up
to about $100\%$ (early MS phase) to $200\%$ (late MS phase) from the
rotation rates obtained in the case of uniform surface fields.
The reduction of the AM loss rate entails a quicker increase of the
stellar rotation rate during the PMS spin-up and a moderated spin-down
during the MS evolution.
The higher the (initial) rotation rate, the larger the AM loss due to 
magnetised winds, and the larger the susceptibility to variations in 
the open flux distribution.
The influence on the rotational history is a cumulative effect, which
depends on the evolutionary stage of the star but eventually also on
the applied model assumptions.

\subsection{Model considerations}
In our wind model the poloidal magnetic field component is prescribed
to be radial \citep[cf.][]{2005smaamldtlmw}, whereas a fully consistent
treatment of the multi-dimensional problem, including the trans-field
component of the equation of motion, is found to show a collimation of
open field lines toward the rotation axis with increasing distance from
the star \citep[e.g.,][]{1985A&A...152..121S}.
The influence of this effect on our results is difficult to quantify,
because investigations focused on this phenomenon are usually limited 
to particular and illustrative cases.
Another approach to include a non-uniform magnetic field topology is
the stellar wind model of \citet{1987MNRAS.226...57M}, which
incorporates in the vicinity of the stellar surface polar `wind' and
equatorial `dead zones', the latter preventing plasma escaping from the
star and thus reducing both the mass and AM loss rate.
Whereas in this model the extent of the dead zone depends on the
rotation rate of the star, the magnetic field distribution at the
surface is prescribed to be dipolar.
Based on the formalism of \citet{1968MNRAS.138..359M} and
\citet{1987MNRAS.226...57M}, \citet{1988ApJ...333..236K} derived a
parametrised description of stellar AM loss rates through stellar
winds, which also comprises qualitative variations of the magnetic
field topology; a definite association of non-uniform magnetic flux
distributions with the respective model parameter is however missing.
Although it remains to be investigated how more complex flux
distributions alter the magnetic field topology of stellar winds
including effects like collimation or dead zones, we deem these aspects
to be less crucial since our analyses are based on \emph{relative}
deviations between two rotational histories.
However, more detailed studies concerning this hypothesis are required.

The surface averaged magnetic field strengths, covering here roughly
two orders of magnitude over the range of relevant rotation rates, are
consistent with observations both in the case of the Sun and rapidly
rotating stars \citep[e.g.,][]{1997MNRAS.291....1D}.
The localised peak field strengths are, depending on the location of
the $50\%$-open flux level and the non-uniformity of the flux
distribution, even larger (up to the order of kilo-Gauss), which is
also in agreement with observations.

The even flux distribution of the Coronal Hole model and the peaked
distribution of the Latitudinal Belt model simulate complementary
non-uniformities and allow us to assess the sensitivity of the results
beyond our $50\%$-open flux classification.
Whereas both cases are consistent within $\lesssim 15\%$, we expect
this tolerance level to be larger for more complex flux distributions.
But in view of the current observational and model limitations a more
sophisticated classification scheme appears yet to be inappropriate.

The thermal wind properties are described by solar-like values, which
are taken to be independent of the rotation rate, the stellar latitude,
and the evolutionary stage of the star.
Although there are indications for a rotation-dependence of the coronal
temperature and density \citep[e.g.,][]{1991MNRAS.252P..21J,
2003ApJ...599..516I}, it is difficult to derive accurate wind
parameters, since observational wind signatures are veiled by the
outshining coronal emission.
The polytropic index of $\Gamma= 1.15$ implies an efficient heating and
thermal driving of the stellar wind; its value lies well in between
those of similar studies \citep[e.g.,][]{1985A&A...152..121S,
2000ApJ...530.1036K}.
It has been chosen to ensure stationary wind solutions at all latitudes
even for small rotation rates, when the magneto-centrifugal driving is
inherently small.
This implies that the contribution of the latitude-\emph{in}dependent
thermal driving is relatively large compared with the
latitude-dependent magneto-centrifugal driving.
In case of a weaker thermal driving, either because of cooler coronae
or a smaller energy flux in the wind, the impact of non-uniform flux
distributions is expected to be even stronger than described above.
If, in turn, magnetised winds turn out to be hotter, latitude-dependent
effects may be diminished (given that the thermal wind properties 
themselves are not latitude-dependent).

The rapidly rotating stellar core represents an AM reservoir which is
tapped by the envelope during the MS evolution.
The efficiency of the internal coupling, quantified through a
characteristic timescale (here determined to be $\tau\dw{c}=
15\un{Myr}$), is under debate, with previous studies investigating
values over a very large range, from very short (solid-body rotation;
e.g., \citealt{1997A&A...326.1023B}) to very long (essentially
decoupled differential rotation; \citealt{1993MNRAS.261..766J,
1998A&A...333..629A}) timescales.
The results are typically not fully consistent with all observational
constraints given by the rotational distributions of stars in young
open clusters of different age, which may point out possible
deficiencies of previous models.
In this respect the non-uniformity of surface magnetic fields presents
an additional and so far rather ignored `degree of freedom', which may
help to match theoretical and observed rotational histories of stars.

\subsection{Relevance to dynamo efficiency and saturation}
Our comparison of the impact of non-uniform flux distributions with the
influence of other magnetic model parameters underlines its importance.
The dynamo efficiency, that is the dependence of the (here, open) flux 
generation on stellar rotation, is generally accepted to increase with 
the rotation rate.
Analytical relations based on the theory of $\alpha\Omega$-dynamos 
(e.g., between the dynamo number and the Rossby number; 
\citealt{1984ApJ...279..763N, 2001MNRAS.326..877M}) as well as 
empirical relationships between the rotation rate and magnetic activity
signatures like the coronal EUV and X-ray emission
\citep{1995A&A...300..775M, 1995A&A...294..515H, 2003A&A...397..147P}
imply power law-dependencies over a large range of rotation rates.
\citet{1972ApJ...171..565S} found the evolution of chromospheric 
activity to decrease with time following $\Omega\propto t^{-1/2}$. 
This relation implies a linear increase of the characteristic stellar
magnetic field strength with rotation rate; an example for the commonly
agreed concept that the decrease of magnetic activity with age reflects
the evolution of the stellar AM and rotation rate.
According to \citet{1991LNP...380..389S}, it is however more the 
magnetic flux which increases linearly, instead of the field strength.
The linear dynamo efficiency is widely used in studies about the 
rotational evolution of stars \citep[e.g.,][]{1995A&A...294..469K,
1997ApJ...480..303K}.
Here, for deviations from the linearity between $\pm30\%$ the rotation
rates on the early MS are found to differ between $150\%$ and $-50\%$.
Our analysis has shown that non-uniform flux distributions can easily 
produce similar deviations. 
A large fraction of open flux in the vicinity of the stellar poles can
therefore compensate the influence of a super-linear dynamo efficiency,
feigning a less efficient (e.g., linear) dependence on the rotation
rate.

A concentration of magnetic flux at high latitudes has qualitatively a
similar influence on the rotational history as a dynamo saturation. 
Since both effects entail a reduction of the AM loss rate, their
synergistic action is expected to enable even higher rotation rates.
Figure \ref{scdscmp.fig} allows for estimates (based on the comparison 
of \emph{exclusive} effects) of the principal contribution to the 
overall reduction of the AM loss.
If the saturation limit of the chromospheric and/or coronal emission
($P\dw{sat,UV/X}\sim 3\ldots1.5\un{d}$) is representative of dynamo
saturation, then we expect non-uniform flux distributions to contribute
only marginally to the reduction of the AM loss.
If, however, the photometric saturation limit ($P\dw{sat,phot}\lesssim
12\un{h}$) reflects the saturation of the dynamo mechanism, then the
rotational evolution of most stars is dominated by the actual
non-uniformity of the surface flux distribution.
But the saturation of indirect activity signatures does not a priori
indicate a saturation of the underlying dynamo processes in the
convective envelope, since the former also depend on, for example,
atmospheric properties and radiation processes which follow different
functional behaviours \citep[e.g.,][]{1997A&A...321..177U}.

\citet{1997A&A...325.1039S} applied a bi-modal flux distribution, 
characterised through polar flux concentrations, to the rotational 
evolution of stars.
For somewhat different model parameters than ours they found a
resemblance between the rotational histories subject to their coronal
hole model and a dynamo saturation limit of $20\un{\Omega_\odot}$,
respectively.
Since this saturation limit is of the order required to explain the
presence of rapidly rotating stars ($\sim 10-20\un{\Omega_\odot}$),
they question the necessity of a dynamo saturation at low rotation
rates and argue in favour for values $\gtrsim 50\un{\Omega_\odot}$.
Whereas our investigation confirms their results in principle, we find
that reductions of the AM loss rate resulting from high-latitude flux
concentrations are smaller; according to our model a flux distribution
with, for example, $\theta_{50\%}\approx 15\degr$ is equivalent to a
dynamo saturation limit of about $40\un{\Omega_\odot}$.
The difference with respect to the \citeauthor{1997A&A...325.1039S}
value ($\sim 20\un{\Omega_\odot}$) may be due to different model
assumptions and parameters.
Since our value is still below their supposed dynamo saturation limit
of $\sim 50\un{\Omega_\odot}$, a strict concentration of magnetic flux
around the poles would be able to dominate the rotational evolution of
stars.
Observations indicate, however, that the $50\%$-open flux level is 
likely located at somewhat lower latitudes.
Surface brightness maps occasionally show and dark elongated features
reaching down to intermediate latitudes \citep[ and references
therein]{2002AN....323..309S}, which shift the mean $50\%$-flux level
(averaged over longitude and evolutionary timescales) equatorwards.
But the presence of magnetic flux in the form of dark spots is a priori
not equivalent with the presence of open magnetic field lines.
A more striking constraint arises from ZDI observations, which in
combination with field extrapolation techniques allow for the
reconstruction of the magnetic field topology and the determination of
the actual surface and latitudinal distribution of open magnetic
fields.
Although the number of detailed observations is yet rather small, first
results show that the principal part of (detectable) open flux is
apparently located at intermediate latitudes
\citep{2004MNRAS.355.1066M}, which motivated our Latitudinal Belt
model.
The cumulated flux distributions of \object{AB Dor} (cf.\ Fig.\
\ref{cuflobs.fig}) show that the average $50\%$-flux level is indeed
located between $30\degr\lesssim \theta_{50\%}\lesssim 60\degr$.
In this range of values our equivalent saturation limit for rapid
rotators is $\gtrsim 60\un{\Omega_\odot}$ and the impact of the
non-uniform flux distribution consequently smaller than the influence
of the dynamo saturation at $\sim 50\un{\Omega_\odot}$ supposed by
\citet{1997A&A...325.1039S}.
In this sense, we consider the effect of non-uniform flux distributions
more as a complementary rather than an alternative mechanism for the
formation of rapid rotators, which is however based on observationally
verifiable principles.

\subsection{Observational constraints}
Compared to the large number of observed surface brightness
distributions of rapidly rotating stars \citep{2002AN....323..309S},
ZDI observations of actual magnetic flux distributions are yet rather
sparse.
A larger database in this field is required to put tighter constraints
on possible open flux distributions, that is the qualitative and
quantitative description of its non-uniformity and dependence on the
stellar rotation rate and the evolutionary phase of the star (including
its dependence on stellar mass).
Whereas, for example, brightness distributions indicate a clear
poleward displacement of starspots with increasing rotation rate, for
open flux a similar relationship is yet not available.
ZDI observations are handicapped through insufficient signal levels
from dark surface regions, so that an unambiguous determination of the
flux topology inside dark polar caps (i.e., whether it is mainly
unipolar and open or more multi-polar and closed) is yet hardly
possible \citep{1997A&A...326.1135D, 2003MNRAS.345..601M}.
If the field reconstructions in the case of \object{AB Dor} and
\object{LQ Hya} prove to be representative and characteristic for rapid
rotators, then we consider the contributions from high-latitude regions
to the total AM loss rate to be of minor importance for the rotational
evolution of rapid rotators.

Owing to their qualitatively similar behaviour it is questionable
whether the impact of non-uniform flux distributions can be
observationally separated from the influence of a dynamo saturation.
This potentially limits the usefulness of rotation rate distributions
as tests of dynamo theories, unless they are supplemented by more
detailed information about the structure of stellar magnetic fields.
A larger observational database is required to refine the rotation rate
beyond which the location of the $50\%$-open flux level is high enough
to cause a discernible impact on rotational distributions of cluster
stars.
Supposed that the limiting rotation rates for high-latitude $50\%$-flux
levels and dynamo saturation are sufficiently low and high,
respectively, it may be possible to distinguish characteristic
signatures of non-uniform flux distributions in the (differential)
distribution of intermediate rotators with rotation rates in the domain
constrained by the two limiting values.

An important role may fall to theoretical models concerning the
pre-eruptive evolution and post-eruptive transport of magnetic flux to
high latitudes in rapidly rotating young stars
\citep[e.g.,][]{2000A&A...355.1087G, 2004MNRAS.354..737M}, since they
allow for a verification of our insight into stellar magnetic
properties by means of activity proxies like (the rotational modulation
of) coronal X-ray and chromospheric UV emission
\citep{1993A&A...278..467V}.
Observations in the UV/EUV spectral range may also be used to refine 
our assumptions about the thermal wind properties, comprising its
latitudinal and rotational dependence, because the total AM strongly
relies on the stellar mass loss rate and on the acceleration of the 
wind through thermal driving.

\section{Conclusion}
\label{conc}
Non-uniform magnetic flux distributions have a significant impact on
the rotational evolution of stars, mimicking the effect of a large
range of dynamo efficiencies and saturation limits.
They present an additional degree of freedom in the modelling of stellar
rotational histories, which can generate differences cumulating up to
$200\%$.
Neglecting their effect implies considerable uncertainties in other
magnetic field-related model parameters such as the dynamo efficiency
of up to about $40\%$.
Although our results are, in principle, in agreement with those of
\citet{1997A&A...325.1039S}, we find the effect of non-uniform flux
distributions to be less efficient than in their investigation; whereas
they find that a concentration of magnetic flux at polar latitudes
imitates a nominal dynamo saturation limit at $20\Omega_\odot$, we find
values of about $35\Omega_\odot$.

Non-uniformities in the form of strong flux concentrations at high
latitudes efficiently reduce the AM loss through magnetised winds,
entailing high stellar rotation rates.
Since magnetic field distributions reconstructed on the basis of ZDI
observations however indicate a considerable amount of open flux at
intermediate latitudes, the anticipated reduction of the AM loss is
expected to be smaller than implied by frequent DI observations of
high-latitude starspots and polar caps.
The influence of non-uniform flux distributions alone thus appears to
be insufficient to explain the existence of very rapid rotators, but
their significant moderation of the AM loss rate makes the requirements
for a dynamo saturation less stringent, enabling saturation limits
$\gtrsim 40\un{\Omega_\odot}$.

\begin{acknowledgements}
The authors are grateful for the referee's suggestions, which improved
the clarity of the paper.
VH acknowledges financial support for this research through a PPARC 
standard grand (PPA/G/S/2001/00144).
\end{acknowledgements}

\bibliographystyle{aa}
\bibliography{}

\end{document}